\documentclass{aastex}
\usepackage{spr-astr-addons}
\usepackage{url}
\usepackage{amsmath}
\usepackage{amsfonts}
\usepackage{amssymb}
\usepackage{graphicx}
\usepackage{subfigure}

\RequirePackage{color}

\def\be{\begin{equation}}
  \def\ee{\end{equation}}
\def\bea{\begin{eqnarray}}
\def\eea{\end{eqnarray}}

\def\n{\nonumber}
\def\l{\label}

\begin{document}

\title{ Modified gravity in a viscous and non-isotropic background}
\shorttitle{ Modify gravity in a viscous and non-isotropic background}

\shortauthors{Kh. Saaidi et al.}
\author{Kh. Saaidi\altaffilmark{1}} \and \author{ A. Aghamohammadi\altaffilmark{2}}
\and  \author{  H. Hossienkhani\altaffilmark{3}}
 \affil{Faculty of Science,  Islamic Azad University Sanandaj Branch, Sanandaj, Iran}
 \altaffiltext{1}{ksaaidi@uok.ac.ir.}
 \altaffiltext{2}{a.aghamohamadi@iausdj.ac.ir}
 \altaffiltext{3}{Hossienhossienkhani@yahoo.com}

\begin{abstract}
 We study the dynamical evolution of an $f(R)$ model of gravity in a viscous and anisotropic background which is given by a Bianchi
type-I model of  the Universe.  We find  viable forms of  $f(R)$  gravity  in which one  is exactly the Einsteinian model of gravity with a cosmological  constant and other  two  are power law $f(R)$ models. We show that these two power law  models  are stable with a suitable choice of parameters. We also examine three potentials which exhibit the potential effect of $f(R)$ models in the context of scalar tensor theory.   By solving  different aspects of the model and  finding the physical quantities in the Jordan frame, we  show that the equation of state parameter satisfy the  dominant energy condition. At last we show that the two power law  $f(R)$ models  behave like  quintessence model at late times and also the  shear coefficient viscosity tends to zero at late times.
\end{abstract}

\section{Introduction}
Observational  data \citep{1,1',2,3} indicates  several shortcoming in the   standard model of gravity (SGR) \citep{4,5,6}. These shortcoming are     related to cosmology, large scale  structure and quantum field theory.
 These problems have spanned  several theoretical models. For example: the quintessence
scenario, which generalizes the cosmological constant
approach \citep{e4, e5}; higher dimensional scenarios \citep{e6,e7, 15}; or the
resort to cosmological fluids with exotic equations of state
\citep{e8,14, 14'}. Another interesting approach  is $f(R)$ gravity, which generalizes the geometrical part of the  Hilbert-Einstein
Lagrangian \citep{c3,c4,c6,f10,p, r1, 9n, ka, ali, k1,  karam1, karam2, la1, la2,r2, r3, r4, r5, NR,16}.

The isotropy of the cosmic microwave background (CMB) radiation was first reported  by the cosmic background
explorer (COBE) satellite \citep{sm}, and subsequently  reinforced by the Wilkinson Microwave Anisotropy
Probe (WMAP) data \citep{Hi}. These observations, coupled with the assumption that we cannot be in a special position in the universe, imply that we live in a   homogeneous and isotropic Universe described by a Friedmann-Lema$\hat{\rm i}$tre-Robertson-Walker (FLRW) line-element. For this reason, nearly all research on the dynamical evolution of the Universe is done in the homogenous and isotropic space-time background.

  On the other hand, although the Universe
seems isotropic and  homogeneous  at present, the large scale matter distribution
in the observable Universe, largely manifested in the form of discrete structures, does
not show homogeneity of a higher order. Moreover,  according to a statistical analysis of 4-yr data
from the COBE satellite,   tiny deviation from  isotropy at the level of $10^{-5}$ has
been suggested  by \citep{20}, and  this suggestion was  confirmed by high resolution WMAP data. These  studies,
 which support the existence of anisotropic phase, lead us to consider the dynamical evolution  of the  Universe with an  anisotropic background.

 A Bianchi type-I (BI) Universe, being the straightforward generalization of the flat
FLRW Universe, is of  interest because it is one
of the simplest models of a non-isotropic Universe  exhibiting   a homogeneity and
spatial flatness. In this case,  unlike the FLRW Universe which has the same scale factor for  three spatial directions, a BI Universe has a different scale factor for each direction. This fact  introduce a non-isotropy to the system. The possible effects of anisotropy in
the early Universe have been  investigated with  Bianchi I type  models from different points of view \citep{ko, ku, Y2, Y3,17,18,19}.
Some other  papers dealing with
anisotropic cosmology are \citep{21,22,23,24,25, 26,27} but, to the best of our knowledge, there are  no  detailed studies
 of    $f(R)$ gravity  in an anisotropic  background. Therefore, we investigate the  dynamical evolution  of the  Universe with a viscous and  anisotropic background in $f(R)$  gravity.

  The present paper is organized  as follows: In  Sec.\;2,  we  review $f(R)$ model and the
 conformal transformation    between  the Jordan frame and the  Einstein frame.
 In  Sec.\;3,  we consider   the
field equations of  $f(R)$   gravity in a non-isotropic space-time.  Sec.\;4,   considers  the set of solutions to the   field
equations. Finally,  the latter section  is  devoted  to
conclusions.
\section{Preliminary}
We study  the Bianchi type I (BI) cosmological model as a gravitational field in modified gravity. The  BI model is
the simplest model of  a non-isotropic Universe that describes a homogeneous and spatially flat
space-time.
\subsection {$f(R)$  gravity }
 We consider the general form of  the action for $f(R)$  gravity to be given by
\begin{equation}\label{1}
S_{f(R)}=\int d^4x\sqrt{-g}\left[\frac{f(R)}{2}+\kappa { L}_{m}(g_{\mu\nu,\psi})\right],
\end{equation}
where $f(R)$ is a function of the Ricci scalar $R$,   $g_{\mu\nu}$ is the metric of space-time,  $\psi$ is the matter field,  and ${ L}_{m}(g_{\mu\nu},\psi)$ is the matter Lagrangian.  We assume  the metric  $g_{\mu\nu}$ is the  only independent variable. Taking the   variation of (\ref{1}) with respect to  $g_{\mu\nu}$ gives
\begin{equation}\label{2}
R_{\mu\nu}f'(R)-\frac{1}{2} f(R)g_{\mu\nu}-\nabla_{\mu}\nabla_{\nu}f'(R)+g_{\mu\nu}\square f'(R)=T^{m}_{\mu\nu},
\end{equation}
where $T^m_{\mu\nu}$, the stress-energy tensor of matter, is  defined by
\be\l{2'}
T^m_{\mu\nu}=-\frac{2}{\sqrt{-g}}\frac{\delta(\sqrt{-g}{
L}_m)}{\delta g^{\mu\nu}}.
\ee
We have taken $\kappa =1$ in this work.
\subsection {Conformal transformation}
As  regards $f(R)$ model of gravity  can be rewritten  as  scalar tensor theories via a famous conformal transformation:
\begin{equation}\label{3}
e^{-2\beta \phi}=f'(R).
\end{equation}
Using   Eq.\;(\ref{3}),  one  can rewrite Eq.\;(\ref{1}) as
\begin{eqnarray}\label{4}
S_{ST}&=&\int d^4x\sqrt{-\bar{g}}\Bigg[\frac{\bar{R}}{2}-\frac{1}{2}\bar{g}^{\mu\nu} \nabla_{\mu}\phi \nabla_{\nu}\phi-V(\phi)\\ \nonumber
&&\;\;\;\;\;\;\;\;\;\;\;\;\;\;\;\;\;\;\;\;\;\;\;  +  {L}_m(e^{2\beta \phi}\overline{g}_{\mu\nu},\psi )\Bigg] ,
\end{eqnarray}
where we have defined  the Einstein frame metric,  $\bar{g}_{\mu\nu}$,   by a conformal transformation $\bar{g}_{\mu\nu}=e^{-2\beta \phi}g_{\mu\nu}$.  The potential $V(\phi)$ is given by
\begin{equation}\label{5}
V(\phi)=\frac{Rf'(R)-f(R)}{2f'(R)^{2}}.
\end{equation}
Variation  of Eq.\;(\ref{4}) with respect to  $\bar{g}_{\mu\nu}$ and $\phi$ gives
\begin{equation}\label{6}
\bar{R}_{\mu\nu}-\frac{1}{2} \bar{R}\bar{g}_{\mu\nu}=\nabla_{\mu}\phi\nabla_{\nu}\phi+\bar{g}_{\mu\nu} \biggr [ \frac{1}{2} (\nabla\phi)^2+V(\phi)  \biggl ]+\bar{T}_{\mu\nu}^m,
\end{equation}
and
\begin{equation}\label{7}
\bar{\square}\phi=V'(\phi)-\beta \bar{T}^{m},
\end{equation}
where $\bar{\nabla}_{\mu}\bar{g}_{\mu\nu}=0$ is used. Note that in the Einstein frame  all indices are lowered and raised by $\bar{g}_{\mu\nu}$.

The line element of the BI type  can be expressed as
\begin{equation}\label{8}
ds^2=dt^{2}-A^{2}(t)dx^{2}-B^{2}(t)dy^{2}-C^{2}(t)dz^{2},
\end{equation}
where the metric function, $A, B, C$, are  functions of time, $t$,
only. This model is an anisotropic generalization of the Friedmann
model
 with Euclidean spatial geometry. The expansion factors, $A$, $B$, $C$, are
 determined via Einstein's equation. In this work,
we shall restrict ourselves to   the Kasner form of the  metric as:
\begin{equation}\label{9}
ds^2=dt^2-t^{2p_{1}}dx^2-t^{2p_{2}}dy^2-t^{2p_{3}}dz^2,
\end{equation}
where $p_{1}$, $p_{2}$, $p_{3}$ are three parameters which we will
 require to be constant. The space is anisotropic if  $p_i\neq p_j ,  (i, j = 1,2,3)$.
\section{Field equations }

By making use of Eqs.\;(\ref{6}) and  (\ref{7}), we can derive the field  equations. Using  Eq.\;(\ref{9})
 and introducing   symbols
 $S\equiv p_{1}+p_2+p_3$ and $Q\equiv p_{1}^2+p_2^2+p_3^2$,
 the components of Ricci  tensor, namely  $\bar{R}_{00}$ and $\bar{R}_{ii}$ are
\begin{equation}\label{10}
\bar{R}_{00}=\frac{S-Q}{t^2}+\frac{S\beta\dot{\phi}}{t}+3\beta\ddot{\phi},
\end{equation}
\begin{equation}\label{11}
\bar{R}_{ii}=t^{2p_i}\left[ \frac{p_i (S-1)}{t^2}-\frac{\beta\dot{\phi}}{t}(S+2p_i)+2\beta^{2}\dot{\phi^{2}}-\beta\ddot{\phi}\right].
\end{equation}
We consider the stress-energy momentum tensor for a viscous fluid:
\begin{equation}\label{12}
T_{\mu\nu}=\left[ \rho+(P-\xi \theta)\right] u_\mu u_\nu-(P-\xi \theta)g_{\mu\nu}+ 2 \eta \sigma_{\mu\nu},
\end{equation}
where $u_{\mu}$, $\rho$, $P$, $\xi$ and $\eta$ are the fluid's four velocity,  energy density,  isotropic pressure,  bulk and shear viscosities, respectively. The scalar expansion and traceless shear tensors are
\begin{equation}\label{13}
\theta_{\mu\nu}=\frac{1}{2}\left [ u_{\mu;\alpha}h^{\alpha}_{\nu}+u_{\nu;\alpha}h^{\alpha}_{\mu}\right ],
\end{equation}
\begin{equation}\label{14}
\sigma_{\mu\nu}=\theta_{\mu\nu}-\frac{1}{3}h_{\mu\nu}\theta,
\end{equation}
respectively. Here,  $h_{\mu\nu}=u_{\mu}u_{\nu}-g_{\mu\nu}$ is the  projection  operator.
Taking the trace of Eq.(\ref{6}) gives :
\begin{equation}\label{15}
\bar{R}=(\bar{\nabla}\phi)^2 +4V(\phi)-\bar{T}^{m}.
\end{equation}
By substituting  Eq.\;(\ref{15}) into  Eq.\;(\ref{6}), we have
\begin{equation}\label{16}
\bar{R}_{\mu\nu}=e^{2\beta\phi}\big[T_{\mu\nu}-\frac{1}{2}Tg_{\mu\nu}\big]+e^{-2\beta\phi}g_{\mu\nu}V(\phi)+\nabla_{\mu}\phi\nabla_{\nu}\phi.
\end{equation}
So the components $\bar{R}_{00}$, $\bar{R}_{ii}$  are
\begin{equation}\label{17}
\bar{R}_{00}=e^{2\beta\phi}\biggr[\rho-\frac{1}{2}\big\{\rho-3(P-\xi \theta)\big\}\biggl]+e^{-2\beta\phi}V(\phi)+\dot{\phi}^{2},
\end{equation}
\bea\label{18}
\bar{R}_{ii}&=&\bigg[ \frac{e^{2\beta\phi}}{2}\left\{ \rho-(P-\xi \theta)+\frac{4\eta}{t}(\frac{S}{3}-p_i)\right\}\\ \nonumber
&&\;\;\;\;\;\;\;\;\;\;-e^{-2\beta\phi}V(\phi) \bigg] t^{2p_{i}}.
\eea
Using Eqs.\;(\ref{10}),  (\ref{11}),  (\ref{17}) and (\ref{18}),  one can gets that

\begin{eqnarray}
&&e^{2\beta\phi}\bigg[\rho-\frac{1}{2}\big\{\rho-3(P-\xi \theta)\big\}\bigg]+e^{-2\beta\phi}V(\phi)+\dot{\phi}^{2}\n \\  &&\;\;\;\;\;\;\;\;\;\;\;\;\;\;=\frac{S-Q}{t^2}+\frac{S\beta\dot{\phi}}{t}+3\beta\ddot{\phi},\label{19}
\eea
\bea
&&p_{i}\bigg[{-2\eta}{t}e^{2\beta\phi}+\frac{ (1-S)}{t^2}+\frac{2\beta\dot{\phi}}{t}\bigg]  +\frac{S\beta\dot{\phi}}{t}-2\beta^{2}\dot{\phi^{2}} +\beta\ddot{\phi}\n \\&&\;\;\;\;\; =
-\frac{e^{2\beta\phi}}{2}\left[ \rho-(P-\xi \theta)+\frac{4\eta S}{3t}\right] +e^{-2\beta\phi}V(\phi)\label{20}.
\end{eqnarray}
Since the  $p_i{\rm s}$ $(i=1,2,3)$ are  linearly independent, we find that
\begin{equation}\label{21}
\eta=\big[\frac{1-S}{2t}+\beta\dot{\phi}\big]e^{-2\beta\phi}.
\end{equation}
Substituting Eq.\;(\ref{21}) into the right hand side of Eq.\;(\ref{20}) gives
\bea\label{22}
&&e^{2\beta\phi}\Big[ \rho-(P-\xi\theta) \Big] =4\beta^2 \dot{\phi}^2-2\beta\ddot{\phi}\\ \nonumber &&\;\;\;\;\;\;\;  -\frac{10}{3}
\frac{S\beta \dot{\phi}}{t}-\frac{2S(1-S)}{3t^2}+2e^{-2\beta\phi}V(\phi).
\eea
By combining Eqs.\;(\ref{22}) and  (\ref{19}), we obtain explicit expression for $\rho$ and $p$ as follows
\bea\label{23}
e^{2\beta\phi}( P-\xi\theta)&=&\frac{1}{6t^2}\left[4S-3Q-{S^2}\right]+\frac{4}{3}\frac{S\beta}{t}\dot{\phi} \\&&\;\;\;\;\;+2\beta\ddot{\phi}
-[\beta^2+\frac{1}{2}]\dot{\phi}^2-e^{-2\beta\phi}V(\phi)\n,
\eea
\begin{equation}\label{24}
e^{2\beta\phi}\rho=\frac{1}{2t^2}(S^2-Q)+e^{-2\beta\phi}V(\phi)-2\beta\frac{S}{t}\dot{\phi}+(3\beta^2-\frac{1}{2})\dot{\phi}^2.
\end{equation}
Substituting  Eqs.\;(\ref{23}) and (\ref{24}) into Eq.\;(\ref{7}), we get that the equations of motion for the scalar field $\phi$, are
\begin{equation}\label{25}
\ddot{\phi}+\dot{\phi}\frac{S}{t}-\beta\dot{\phi}^{2}=\frac{e^{-2\beta\phi}}{(1-6\beta^2)}\left[V'(\phi)-4\beta V(\phi)
\right]+\frac{C}{t^2},
\end{equation}
where $$C=\frac{\beta}{(1-6\beta^2)}\left( 2S-S^2-Q\right).$$
\section{The solutions }

In this section we want to solve Eq.\;(\ref{25}).  As an ansatz, we suggest
\begin{equation}\label{26}
\phi=\phi_0 \ln (t),
\end{equation}
where $\phi_0$ is a constant with the same dimension as $\phi$ and $t$ is a dimensionless time parameter such as $t={\tau/\tau_0}$, where $\tau_0$ can be the present time. By this choice, we can consider three classes  of solution as follows.

\subsection{Case I}
 A class of  solutions to  Eq.\;(\ref{25})  is given by
\begin{eqnarray}
&&\frac{e^{-2\beta\phi}}{(1-6\beta^2)}\Big[V'(\phi)-4\beta V(\phi)\Big]=0,\label{27} \\
&&\ddot{\phi}+\dot{\phi}\frac{S}{t}-\beta\dot{\phi}^{2}
-\frac{C}{t^2}=0.\label{28}
\end{eqnarray}
 In this case we assume $\phi=\phi_1 \ln (t)$, where $\phi_1 $ is a constant. Solving Eq.\;(\ref{27}), gives
\begin{equation}\label{29}
V(\phi)=V_1 e^{4\beta\phi},
\end{equation}
where $V_1$ is an integration constant. By substituting Eq.\;(\ref{29}) into Eq.\;(\ref{28}),  we  obtain  the following  constraint:
\begin{equation}\label{30}
(S-1)\phi_1-\beta\phi_1^2-C=0.
\end{equation}
Also,  we can find that  $\rho$, $P$ and $\eta$ are
\begin{equation}\label{31}
\rho=V_1+\frac{\rho_1}{t^{2\alpha_1}},
\end{equation}
\begin{equation}\label{32}
P=-V_1+\frac{S}{t}\xi+\frac{P_1}{t^{2\alpha_1}},
\end{equation}
and
\begin{equation}\label{33}
\eta=\left[{1-S}+2\beta\phi_1\right]\frac{1}{2t^{2\alpha_1-1}}.
\end{equation}
Where
\be
\rho_1= {1\over 2}\bigg[{S^2-Q}+\biggr \{(6\beta^2-1)\phi_1-2\beta S \biggl\} \phi_1  \bigg],
\ee
\be
P_1={1\over 6}\bigg[4S-3Q-S^2+\left\{  8S\beta-12\beta-(6\beta^2+3)\phi_1\right\}\phi_1\bigg],
\ee
 and
$$\alpha_1 =(\beta\phi_1+1).$$
From Eqs.\;(\ref{31}) and (\ref{32}) we can find the equation of state parameter, $\gamma=P/\rho$ as
\begin{equation}
\gamma= {P \over \rho} = -1 + {(\gamma_1 +1)\rho_1 \over V_1 t^{2\alpha_1} +\rho_1} + {S\xi \over V_0 t + {\rho_0 \over t^{2\alpha_1}-1}},
\end{equation}
here $\gamma_1 = P_1 /\rho_1$.

Using Eqs.\;(\ref{29})  and  (\ref{5}),  one can obtain that the viable  $f(R)$ and $V(R)$ function  are
\bea\label{34}
f(R)&=&-2V_1+f_1R,\\
V(R)&=& {V_1 \over f_1^2}
\eea
where $f_1$ is an integration constant and $2V_1$ is similar to the  cosmological constant $\Lambda$. In this case, the  $f(R)$ model of gravity   is exactly   the standard  Einsteinian model of gravity.
\subsection{Case II}

   We assume $\phi=\phi_2 \ln (t)$, where $\phi_2 $ is a constant. So,  we suggest  another class of solutions which satisfy the following equations:
\begin{eqnarray}
\frac{e^{-2\beta\phi}}{(1-6\beta^2)}\Big[V'(\phi)-4\beta V(\phi)\Big]&=&\frac{B}{t^2},\label{35}\\
\ddot{\phi}+\dot{\phi}\frac{S}{t}-\beta\dot{\phi}^{2}-\frac{B+C}{t^2}&=&0,\label{36}
\end{eqnarray}
where $B$ is a constant. From Eq.\;(\ref{35}), we have
\begin{equation}\label{37}
V(t)=V_2t^{2\beta\phi_2-2},
\end{equation}
where
\begin{equation}\label{38}
V_2=\frac{B(6\beta^2-1)\phi_2}{2(1+\beta\phi_2)}.
\end{equation}
By substituting Eq.\;(\ref{37}) into Eq.\;(\ref{36}), we can obtain a constraint  which   $\phi_2 $ has to   satisfy:
\begin{equation}\label{39}
(S-1)\phi_2-\beta\phi_2^{2}-B-C=0.
\end{equation}
Also,  we get that  $\rho$, $P$ and $\eta$  are
\begin{eqnarray}\label{40}
\rho &=&{\rho_2 \over t^{2(\beta\phi_2 +1)}} + V_0 t^{2(\beta\phi_2-1)} ,\\
P&=&{P_2 \over  t^{2(\beta\phi_2 +1)}} - V_0 t^{2(\beta\phi_2-1)}
+\frac{S}{t}\xi ,\label{41} \\
\eta &=&{1\over 2}\left[1-S+2\beta\phi_2\right]\frac{1}{t^{(2\beta\phi_2+1)}}.\l{42}
\end{eqnarray}
where
\be
\rho_2={1\over 2}\left[{S^2-Q}+ (6\beta^2-1)\phi_2^2-4\beta S \phi_2  \right],
\ee
and
\be
P_2={1\over 6}\bigg[
(8S\beta-12\beta)\phi_2-(6\beta^2+3)\phi_2^2+4S-3Q-S^2\bigg].
\ee
From Eqs.\;(\ref{40} ) and (\ref{41}) we have
\begin{equation}
\gamma= {P \over \rho} = -1 + {P_2+\rho_2 + S\xi t^{(2\beta\phi_2 -1)}\over \rho_2 + V_0t^{4\beta\phi_2}}.
\end{equation}

 Using Eqs.\;(\ref{37}) and (\ref{5}), we get that  the viable  $f(R)$ and $V(R)$ functions are:
\bea\label{43}
f(R)&=& f_2R^{1- \epsilon},\\
V(R)&=&-{\epsilon \over 2(1-\epsilon)^2f_2}R^{1+\epsilon},\\\l{51}
V(\phi) &=& -{\epsilon\big[ (1-\epsilon)f_2\big]^{1\over \epsilon} \over 2(1-\epsilon)^2f_2}  e^{{2\beta(1+\epsilon)\over \epsilon}\phi},\l{52}
\eea
where $f_2$  is the constant of integration  and
\be\l{53}
\epsilon={B\phi_2 \over B\phi_2+ S^2+Q-4S}.
\ee
To obtain stability for the above  $f(R)$,   the following condition  must   be satisfied \citep{ali}:
\begin{equation}\label{46}
\frac{d^2 f}{dR^2}>0.
\end{equation}
One can see that Eq.\;(\ref{43}) is stable when
\begin{equation}\label{47}
\epsilon < 0.
\end{equation}
From Eqs.\;(\ref{43}) and (\ref{52}), one can see that for $\beta\phi_2=0 (\epsilon=0)$, it follows that   $f(R)=R$,  which is the standard Einsteinian model of gravity. However,  for  $\beta\phi_2=0 (\epsilon=0)$,  the quantities  $\rho$, $P$ and $\eta$   are:
\begin{equation}\label{49}
\rho=\frac{S^2-Q}{2t^2},
\end{equation}
\begin{equation}\label{50}
P-\frac{S}{t}\xi=\dfrac{4S-S^2 -3Q}{6t^2},
\end{equation}
\begin{equation}\label{51}
\eta=\frac{1-S}{2t}.
\end{equation}
These results are  in   \citep{26}.  The dominant energy condition (DEC)  requires   $-\rho+\frac{S}{t}\xi\leq P\leq\rho+\frac{S}{t}\xi$. Then, from Eqs.\;(\ref{49}),    (\ref{50}) and (\ref{51}), for  $P\leq\rho+\frac{S}{t}\xi$,  we have that $1\leq S$. For  the  case $-\rho+\frac{S}{t}\xi\leq P$, we obtain  $S(S+2)\geq 3Q$. Thus, from these expressions it is easy to see that  $\eta\leq0$ and $\rho\geq0$.

\subsection{ Case III}
In this case we assume $\phi =\phi_3 \ln(t) $ and  $B=-C$.  According to  Eq.\;(\ref{39})  $\phi_3 $ is
\begin{equation}\label{52}
\phi_3 = {S-1 \over \beta},
\end{equation}
 and Eq.\;(\ref{37}) reduces to
\begin{equation}\label{53}
V(t)=V_3t^{2(S-2)},
\end{equation}
where
$$V_3=\frac{(2S-S^2-Q)(S-1)}{2S}.$$
We can obtain physical quantities as
\begin{eqnarray}\label{54}
\rho &=&{\rho_2 \over t^{2S} }+V_3t^{2(S-2)},\\
P&=&{P_2 \over  t^{2S}}
+\frac{S}{t}\xi-V_3t^{2(S-2)},\label{55} \\
\eta &=&\frac{(S-1)}{2t^{2S-1}}.\label{56}
\end{eqnarray}
Here
\be
\rho_3={1\over 2}\left[{3S^2-Q}-8S+6-\frac{(S-1)^2}{\beta^2} \right],
\ee
\bea
P_3&=&{1 \over 6}\bigg[S^2-4S+6-3Q-\frac{3(S-1)^2}{\beta^2}\bigg],
 \eea
and  we can  obtain  $f(R)$ and $V(R)$  as
\bea\label{57}
f(R)&=& f_3R^{1+\epsilon},\\
V(R) &=& \Big[{W-2V_3 \over 2Wf_3}\Big] R^{1-\epsilon},\\
V(\phi) &=& \Big[{W-2V_3 \over 2W^{\epsilon}f_3}\Big]e^{{2\beta(1-\epsilon)\over S-1}\phi},
\eea
where $f_3$ is an integration constant and
\be
\epsilon = {2V_3 \over W-2V_3} , \hspace{1.5cm} W= {1\over 4} (4S-S^2-Q).
\ee
One can show that this model is stable if $\epsilon >0$.

{Since the shear coefficient of viscosity at early time is much
bigger than the bulk viscosity \citep{27},       we take
$\xi=0$. Moreover,  for simplicity we suppose a special non-isotropic background given  by $p_1=p_2 =p $ and  $p_3= np$, where $n$ is a positive  constant.    However,  one can write the barotropic equation of state as
$P=\gamma\rho$. It is well known that the dominant  energy condition
(DEC) imply that $\gamma$ lies in the interval [-1, 1]. From
Eqs.\;(\ref{54}) and (\ref{55}), and the above assumptions, we can obtain the
 equation of state parameter   as
\begin{equation}\label{58}
\gamma=-1 +{P_3 + \rho_3 \over \rho_3 + V_3 t^{4(S-1)}}.
\end{equation}
In order to gain  better insight we make   a suitable  choice of parameters. For example, we  let $p=0.45$, $n=1.2$ and $\beta=\sqrt{2/3}$.  For this choice,  $S=1.44$ and $\epsilon \approx 0.047  $, which  means that the model is stable. We plot the equation of state parameter $\gamma$ and the
shear coefficient of viscosity $\eta$ versus $t$ if Fig.\;1. Fig.\;1a indicates that the equation of state parameter belongs to  $[-1, 1]$. I.e., this model can satisfy the DEC,  and by increasing time it is  saturated to $-1$. This means  our model behave like a   quintessence model at late times. Fig.\;1b shows that the shear coefficient of viscosity is notable at  early times and as time passes it tends to zero.  Namely,
the part of viscosity  coming from the non-isotropic form of background is negligible at late times ( the present time).}

\begin{center}
\begin{figure}[t]
\begin{minipage}[b]{1\textwidth}
\subfigure[\label{fig1a} ]{ \includegraphics[width=.3\textwidth]%
{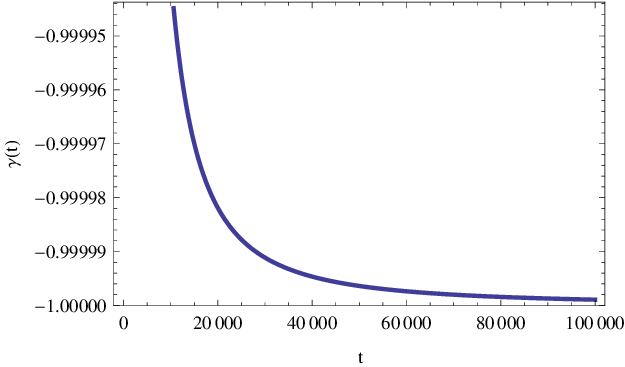}} \hspace{0cm}\\
\subfigure[\label{fig1b} ]{ \includegraphics[width=.3\textwidth]%
{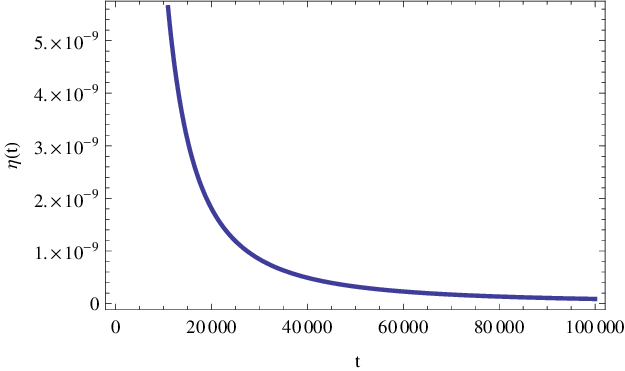}}
\end{minipage}
\caption{ (a): This sub-figure shows the equation of state parameter, $\gamma$, versus cosmic time. (b): This sub-figure shows the shear coefficient viscosity,  $\eta$, versus cosmic time, $t$. We have taken $p=0.45$, $n=1.2$ and $\beta=\sqrt{2/3}$.  }
\end{figure}
\end{center}
\newpage
\section{Conclusion}
 The main purpose of the present work  is to  study  $f(R)$  gravity in an anisotropic metric, such as the Kasner
metric.  We have assumed  that the Universe is filled by a viscous cosmic fluid which is endowed with a shear viscosity $\eta$ and a bulk
viscosity $\xi$. In this work, we have obtained three  viable forms of $f(R)$  gravity in which one  was exactly the Einsteinian model of gravity with a cosmological  constant and the other two  were power law $f(R)$ models. We    showed that these two power law modified models of gravity are stable with a suitable choice of parameters. Also,  we solved  different aspects of the model and  found   physical quantities in the Jordan frame. Our study  shows that the equation of state parameter satisfies the  dominant energy condition. Moreover,   the power law  $f(R)$ models behave like a  quintessence model at late times and the   shear coefficient viscosity tends to zero at late times.

\section{Acknowledgement}
The authors thank the referee for   illuminating
remarks that enabled them to improve the clarity of this  paper and also the authors would like thank to S. Westmoreland for assisting for clarity the English in this paper.

\end{document}